\documentclass[aps,reprint,prc,showpacs,amsmath,amssymb,nofootinbib]{revtex4-1}
\usepackage{graphicx}
\usepackage{color}
\usepackage{times}
\usepackage{inputenc}
\usepackage{bm}
\usepackage{multirow}
\usepackage{float}
\usepackage{url}
\usepackage{natbib}
\usepackage{tensor}
\usepackage[colorlinks=true,citecolor=blue,urlcolor=blue,linkcolor=red]{hyperref}
\usepackage{comment}
\usepackage{stackengine}

\newcommand{\crd}{\textcolor{black}}

\newcommand{\be}{\begin{eqnarray}}
\newcommand{\ee}{\end{eqnarray}}
\newcommand{\ba}{\begin{align}}
\newcommand{\ea}{\end{align}}

\begin{document}
	\hspace{5cm}IP/BBSR/2021-4
	\vspace{0.5cm}
	
	\title{Displaced Neutrino Jets  at the LHeC}
	\author{Giovanna Cottin$^{1}$}
	\email{giovanna.cottin@uai.cl}
	\author{Oliver Fischer $^{2}$}
	\email{oliver.fischer@liverpool.ac.uk}
	\author{Sanjoy Mandal$^{3}$}
	\email{smandal@ific.uv.es}
	\author{Manimala Mitra$^{4,5}$}
	\email{manimala@iopb.res.in}
	\author{Rojalin Padhan$^{4,5}$}
	\email{rojalin.p@iopb.res.in}
	\affiliation{\it $^{1}$ Departamento de Ciencias, Facultad de Artes Liberales,Universidad Adolfo Ib\'a\~{n}ez, Diagonal Las Torres 2640, Santiago, Chile }
	
	\affiliation{\it  $^{2}$ Department of Mathematical Sciences, University of Liverpool, Liverpool, L69 7ZL, UK}
	
	\affiliation{ \it $^{3}$ AHEP Group, Institut de F\'{i}sica Corpuscular --
		CSIC/Universitat de Val\`{e}ncia, Parc Cient\'ific de Paterna.\\
		C/ Catedr\'atico Jos\'e Beltr\'an, 2 E-46980 Paterna (Valencia) - SPAIN }
	\affiliation{\it $^{4}$Institute of Physics, Sachivalaya Marg, Bhubaneswar-751005, India}
	\affiliation{\it $^{5}$Homi Bhabha National Institute, Training School Complex,
		Anushakti Nagar, Mumbai 400094, India}
	\date{\today}

\begin{abstract}
Extending the Standard Model  with right-handed neutrinos (RHNs) is well motivated by the observation of neutrino oscillations.
In the type-I seesaw model,  the RHNs interact with the SM particles via tiny mixings with the active neutrinos, which makes their discovery in the laboratory, and in particular at collider experiments  in general challenging.
In this work we instead consider an extension of the type-I seesaw model  with the addition of  a  leptoquark (LQ), and employ a non-minimal production mechanism of the RHN via  LQ decay, which is  unsuppressed by neutrino mixing. We focus on relatively light RHN with mass $\mathcal{O}(10)$ GeV and LQ with mass 1.0 TeV, and   
explore the discovery prospect of  the RHN at the proposed Large Hadron electron Collider. %
In the considered mass range and with the given interaction strength, the RHN is long lived and, due to it stemming from the LQ  decay, it is also heavily boosted,  resulting in  collimated decay products. The unique signature under investigation is thus  a displaced fat jet. 
We use kinematic variables to separate signal from background, and demonstrate that the ratio variables with respect to energy/number of displaced and prompt tracks are useful handles  in the  identification of displaced decays of the RHN.
We also show that employing a positron beam provides order of magnitude  enhancement in   the detection prospect of this signature. 
\end{abstract}

\maketitle

{ \bf Introduction.--}
The observation of neutrino oscillations is a clear indicator for new physics beyond the Standard Model~(BSM). A plethora of models exist that aim at explaining the light neutrino masses and mixings, many of which contain Standard Model (SM) gauge singlet right-handed neutrino (RHN). 
The simplest one among them is the type-I seesaw model \cite{Minkowski:1977sc}, where the RHNs are Majorana particles.
Being SM gauge singlet RHNs interact with the SM particles only via their mixing with the active neutrinos, referred to as active-sterile mixing, and is proportional to $\sqrt{\frac{m_\nu}{M_N}}$, where $m_{\nu} $ and $M_N$ are the light neutrino and the RHN mass scale. 
Since  $m_\nu <$ eV, this mixing is small, leading to a suppressed  production of RHN at colliders, which makes  their observation  challenging.
This limitation can be avoided if RHNs are embedded in a theory framework with a new production mechanism that involves unsuppressed interactions of RHN with  BSM/SM particles.

Motivated by this, we consider a theory framework that includes a $\tilde{R}_2$ leptoquark (LQ) and Majorana RHN. We consider a mass of $1$ TeV for LQ and focus on RHN with low mass  $M_N \sim \mathcal{O}(10)$ GeV.
 The RHN in this framework interacts with the LQ and an up-type quark and can thus be produced from LQ decay.
For the considered mass scales, the RHN is heavily  boosted and its decay products leptons and jets are collimated, which fail to satisfy standard isolation criteria, thereby enabling a large radius jet description as the appropriate description to adopt. Additionally,  a RHN in the considered mass range is a long lived particle and its decay is displaced from the point of its creation.
\crd{The final state object of interest is thus a displaced large radius jet, or a fat jet, which is also accompanied by a prompt jet.} 

This signature is challenging to probe at the LHC due to its hadronic nature.
Therefore, we investigate this signature in a relatively clean environment, {\it i.e.,} in lepton-hadron collisions at the proposed Large Hadron electron Collider (LHeC)~\cite{AbelleiraFernandez:2012cc}. The LHeC allows for resonant $\tilde R_2$ production with a sizeable cross-section, which has been shown to have a very good sensitivity to a LQ with first-generation coupling~\cite{Zhang:2018fkk}, cf\ also ref.~\cite{Agostini:2020fmq,Padhan:2019dcp}.  
Moreover, the possibilities of electron polarisation and to use a positron beam provide extra handles to test the nature of the new physics signature~\cite{AbelleiraFernandez:2012cc,Padhan:2019dcp}. 

The paper is organized as follows: First we briefly review the model and the existing constraints on LQ. Following that we discuss the production of RHN at an $ep$ collider LHeC. In the subsequent section, we present a detailed collider analysis and discuss discovery prospect of  the unique RHN signature. Finally, we present the summary of the paper.

\begin{figure}[b]
	\centering	
			\includegraphics[width=0.35\textwidth,height=0.2\textheight]{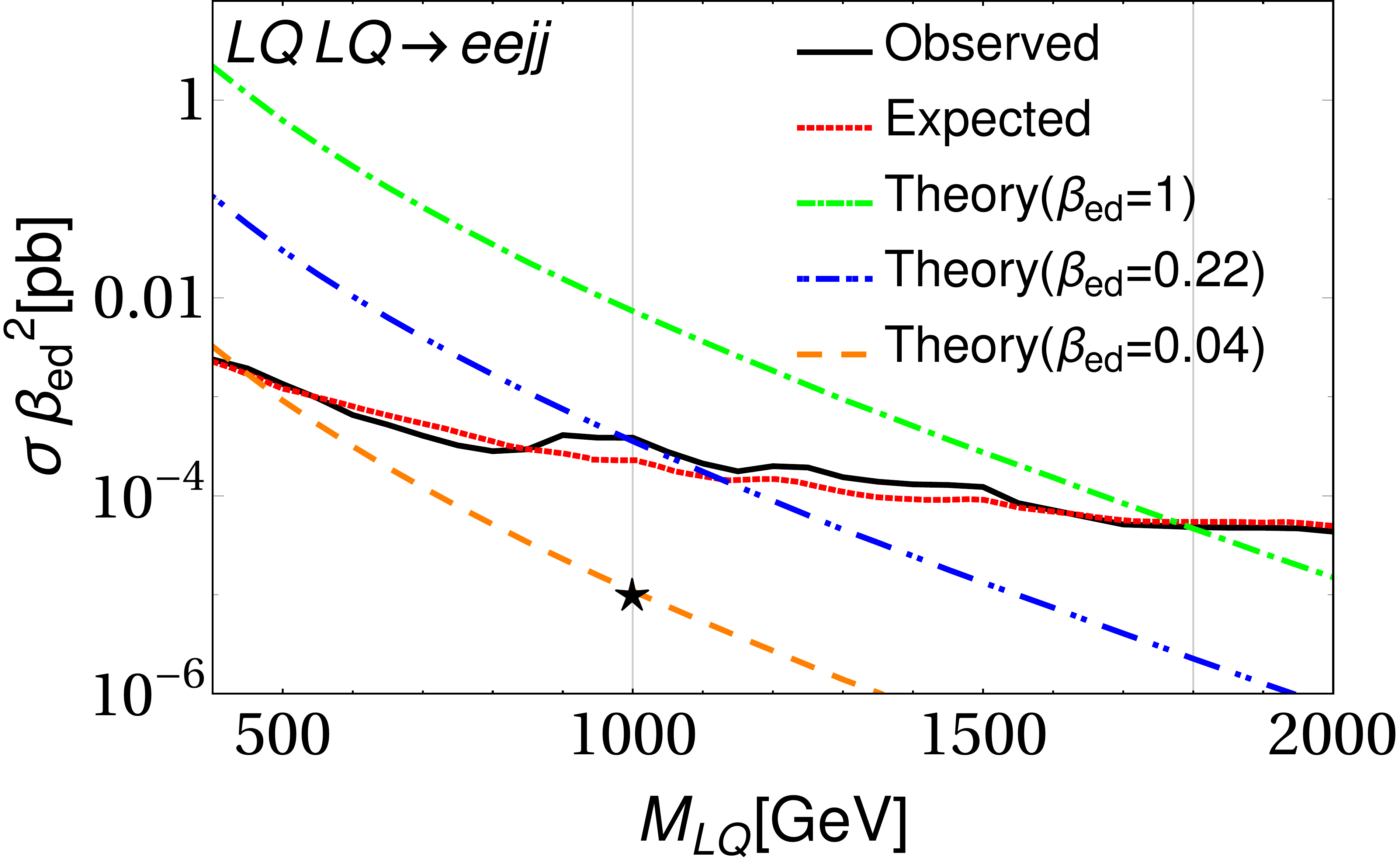}
	\caption{Comparison between theory cross-section and the observed limit on $eejj$ cross-section from the LHC. The black line is the 13 TeV LHC  limit on  $\sigma \beta_{ed}^2$ \cite{Aad:2020iuy}. The green, blue, orange lines represent the variation of theory cross-section $\sigma (p p \to LQ LQ) \times \beta_{ed}^2$ w.r.t LQ mass. The star denotes our chosen benchmark couplings $Y_{11}=0.2, Z_{11}=1$, also consistent with the APV bound.}
	\label{fig:limit13}
\end{figure}

{ \bf The Model.--} 
We consider the $\tilde R_2$ leptoquark model, which contains a  scalar LQ, and three  RHN states ($N_i$). The LQ has two iso-spin components, $\tilde{R}_2(3,2,1/6)=(\tilde{R}_2^{\frac{2}{3}},\tilde{R}_2^{-\frac{1}{3}})^T$, where the superscript of the  components denotes the electromagnetic charge. The  following renormalizable terms  describe interactions of $\tilde{R}_2$ with the SM fermions and $N_i$ ~\cite{Dorsner:2016wpm,Buchmuller:1986zs,Buchmuller:1986iq,Padhan:2019dcp,Mandal:2018qpg}:
\begin{align}
\mathcal{L}_{LQ} = - Y_{ij}\bar{d}_{R}^{i}\tilde{R}_{2}^{a}\epsilon^{ab}L_{L}^{j,b}+
Z_{ij}\bar{Q}_{L}^{i,a}\tilde{R}_{2}^{a}N_{R}^{j}+\text{H.c.}
\label{eq1}
\end{align}
In the above,  $i,j=1,2,3$ are flavor indices, $a,b=1,2$ are $SU(2)_L$ indices, \crd{and $N_R$ indicates the right chiral component of $N$}. Upon expansion, Eq.~\eqref{eq1} becomes:
\begin{align}
&\mathcal{L}_{LQ} =-Y_{ij}\bar{d}_{R}^{i}e_{L}^{j}\tilde{R}_{2}^{2/3}+(YU_{\text{PMNS}})_{ij}\bar{d}_{R}^{i}\nu_{L}^{j}
\tilde{R}_{2}^{-1/3}+\\ \nonumber
& Z_{ij}\bar{u}_{L}^{i}N_{R}^{j}\tilde{R}_{2}^{2/3}+(V_{\text{CKM}}Z)_{ij}\bar{d}_{L}^{i}
N_{R}^{j}\tilde{R}_{2}^{-1/3}+\text{H.c.}.
\label{eq2}
\end{align}
%
Here, $Y$ and $Z$ are the $3\times 3$  complex Yukawa coupling matrices, $V_{\text{CKM}}$ and $U_{\text{PMNS}}$ are the Cabibbo-Kobayashi-Maskawa and Pontecorvo-Maki-Nakagawa-Sakata matrices. 
To investigate the model signature,  it is sufficient to assume that only one generation of RHN couples with leptons and quarks. We consider this to be  $N_1$ (denoted henceforth as $N$), with  only   $Z_{11}\neq 0$, and all other $Z_{ij}$ being 0.  Additionally, we also consider only $Y_{11} \neq 0$.  
Due to the $Y_{11}, Z_{11}$ couplings  $\tilde{R}^{2/3}_2$ can decay to both $ed$ and $Nu$  states. We denote the corresponding branching ratios by $\beta_{ed}$ and $ \beta_{Nu}$, respectively.

{\bf Constraints.--} Both direct and indirect experimental searches  give strong constraints on a  LQ state. 
Very relevant for LQ production at the LHeC is the precision measurement of atomic parity violation~(APV), which tightly constrains the LQ coupling to a $d$ quark and $e$ state, as ${Y_{11}}<0.34\frac{M_{\text{LQ}}}{1\,\text{TeV}}$ ~\cite{Dorsner:2014axa}, $M_{LQ}$ is the LQ mass. Evidently, for larger LQ mass the constraint on  coupling ${Y_{11}}$ relaxes. The tree level lepton flavour violating~(LFV) Kaon decay   $K_L\to\mu^-e^+$ also gives a tight constraint on the  LQ couplings  $|Y_{22}Y_{11}^*|\leq 2.1\times 10^{-5}(\frac{M_{\text{LQ}}}{1\,\text{TeV}})^2$ \cite{Dorsner:2014axa,Dorsner:2011ai}. 
 Due to our choice of  $Y_{22}$ = 0, this is not relevant for our study. 

{\bf LHC search.--}
LQ are pair-produced in proton-proton collisions, which has been investigated in few different channels from the process $pp\to \text{LQ}\, \text{LQ}\to \ell j\ell j$~\cite{Sirunyan:2018btu,Aad:2020iuy}.
Non-observation of any signal at $\sqrt{s}=13$ TeV constrains LQ masses to be larger than 1.8 TeV at $95\%$ C.L~\cite{Aad:2020iuy} for LQ coupling exclusively to the first generation fermions. Including  several decay channels simultaneously,  in particular $LQ\to N u$, relaxes the constraint on the LQ mass,  as shown in Fig.\ref{fig:limit13}. In the following we consider a large branching ratio $\beta_{Nu}=96\%$ for $\tilde R_2 \to Nu$ production mode, which leaves a sufficiently small branching ratio $\beta_{ed}$ for the  $\tilde{R}_2 \to de$ decay to relax the LHC mass-limit to below 1 TeV. 
These branching ratios can be obtained for $Y_{11}=0.2, Z_{11}=1.0$. The point marked in star in Fig.\ref{fig:limit13} represents this benchmark. 

It is important to realise that, as $\beta_{ed}$ decreases, the branching ratio $\beta_{Nu}$ becomes large, and   $LQ \to Nu$ process could in principle be investigated at the LHC. We remind ourselves that the decay $\tilde R_2 \to N u$ is followed by the decay of $N$ leading to a displaced fat jet for  $M_N={\cal O}(10)$ GeV. 
We find that for our benchmark point the cross-sections for $p p \to LQ LQ \to Nu Nu$, and $p p \to LQ LQ \to ed Nu$ are $\sigma=7.53, 0.31$ fb, respectively, which are much smaller than what we obtain from  $e p \to N u$ at LHeC, as discussed below.
Additionally, the final state being all-hadronic and being hidden in the large QCD background at the LHC, we expect the sensitivity to be rather small. 
Hence for the study of the displaced fat jet,  we rather choose to work with LHeC. 

\begin{figure}
	\centering
	\includegraphics[width=0.3860\textwidth,height=0.140\textwidth]{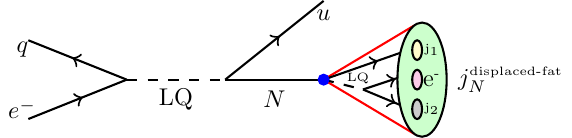}
	\caption{\small{Sample Feynman diagram  for the model signature at the LHeC, the blue circle indicates displaced $N$ decay.}}
	\label{fig:feynmandiag}
\end{figure}

{\bf LQ  decays.--}
For non-zero $Y_{11}, Z_{11}$ couplings $\tilde R_2^{2/3}$ can decay to both $ed$ and $N u$ states.
The analytical expression for these two-body decays are, 
\begin{align}
\Gamma(\text{LQ}\to e d/N u)=\frac{|Y_{11}|^{2}f_1/|Z_{11}|^{2}f_2}{16\pi M_{\text{LQ}}^{3}}\,,
\end{align}
where
$ f_1=\lambda^{\frac{1}{2}}(M_{\text{LQ}}^{2},m_{e}^{2},m_{d}^2)(M_{\text{LQ}}^{2}-m_{e}^{2}-m_{d}^{2}),  f_2=\lambda^{\frac{1}{2}}(M_{\text{LQ}}^{2},M^2_{N},m^2_{u})(M_{\text{LQ}}^{2}-M_{N}^{2}-m_{u}^{2})$ and $\lambda(x,y,z)=x^2+y^2+z^2-2xy-2xz-2yz$.
%
%
As shown by the green line in Fig.~\ref{fig:limit13}, the LHC searches for the decay $LQ \to ed$ constrains  LQ mass as $M_{\textrm{LQ}}>1.80$ TeV, assuming a  branching ratio $\beta_{ed} = 1$.
Considering smaller branching fractions this lower bound relaxes to, for instance, $M_\textrm{LQ} = 1.0$ TeV when $\beta_{ed}=0.22$. For $\beta_{ed} = 0.04$ the constraint is even more relaxed. 

{\bf RHN  decay.--}
\crd{For the  chosen values of the  couplings}, the dominant decays of $N$ are mediated via an off-shell $\tilde R_2$. The contributions of SM mediators are suppressed by  small active-sterile mixing. 
For the  considered mass range $N$ decays to $N \to  e^{-} u  \bar{d}, e^+\bar{u} d$ \crd{via  $\tilde{R}^{2/3}_2$}~(two-body decays into lepton and pseudoscalar/vector meson are relevant for $M_N \leq {\cal O}(1)$ GeV). $N$ can also decay to $N\to\nu q \bar{q}^\prime$ via \crd{ $\tilde{R}^{-1/3}_2$}.  
The  partial decay width for $N \to e^- u \bar{d}/e^+ \bar{u} {d}$  is:
\begin{align}
\Gamma(N \to e^-u \bar{d}/e^+ \bar{u} {d})=N_c\frac{|Z_{11}|^2 |Y_{11}|^2 M_{N}^5}{512 \pi^3 M_{LQ}^4} \mathcal{I} 
\label{N three body decay}
\end{align}
Here, $\mathcal{I}=I(x_{u},x_{d},x_{e})=\int_{(x_{d}+x_{e})^2}^{(1-x_{u})^2}\frac{dz}{z} (1+x_{u}^2-z) (z-x_{d}^2-x_{e}^2)\lambda^{\frac{1}{2}}(1,x_{u}^2,z)\lambda^{\frac{1}{2}}(z,x_{d}^2,x_{e}^2)$ and $x_{u/d/e}=\frac{m_{u/d/e}}{M_{N}}$, 
and  $N_c=3$ is the color factor. $N\to\nu q \bar{q}^\prime$ also has similar dependency on masses and couplings. The branching ratios for $N$ decaying into $ eq \bar{q}^\prime$ and $\nu q \bar{q}^\prime$ are 50\% each. For $M_N < 50$ GeV, $N$ has a proper decay length $c\tau_N > {\cal O}(1)\,\mu m$ for our chosen benchmark point. In particular for $M_N =  10$ GeV (20 GeV),  $c\tau_N \sim 1 \  mm$ (\crd{$c \tau_N \sim 0.01 \  mm$}).
On the other hand, for $M_{N} \geq 50$ GeV,  $c\tau_N $ 
is less than $1\ \mu m$, which is below the resolution of the LHeC and also  LHC detectors, such that $N$ decay cannot be considered as being displaced.

{\bf Signal.--} 
Our stage is the LHeC with its \crd{$7$ TeV  proton and 60 GeV } electron beam and without polarisation, where $N$ is produced via $e p \to  N u$ process. 
The signal is dominated by resonant $\tilde R_2^{2/3}$ production
 with a cross section of 9.3 fb, followed by  the subsequent decay $\tilde R^{2/3}_2 \to N u$ with 96\% branching ratio, see Fig.~\ref{fig:feynmandiag} for sample diagram. \crd{This production channel  strongly depends on the two couplings $Y_{11}$ and $Z_{11}$.} 
Also the $t$-channel contribution, mediated by $\tilde R_2^{2/3}$, contributes sizeably with $\sigma \sim 5.7$ fb.  We remark, that the production of $N$ via leptonic mixing \crd{is suppressed by  mixing  square} $m_\nu/M_N < {\cal O}(10^{-10})$ and is therefore completely negligible in our model. \crd{The associated parton $u$ hadronises and gives rise to a prompt light jet.} 

\begin{figure}
	\centering
	\includegraphics[width=0.40\textwidth]{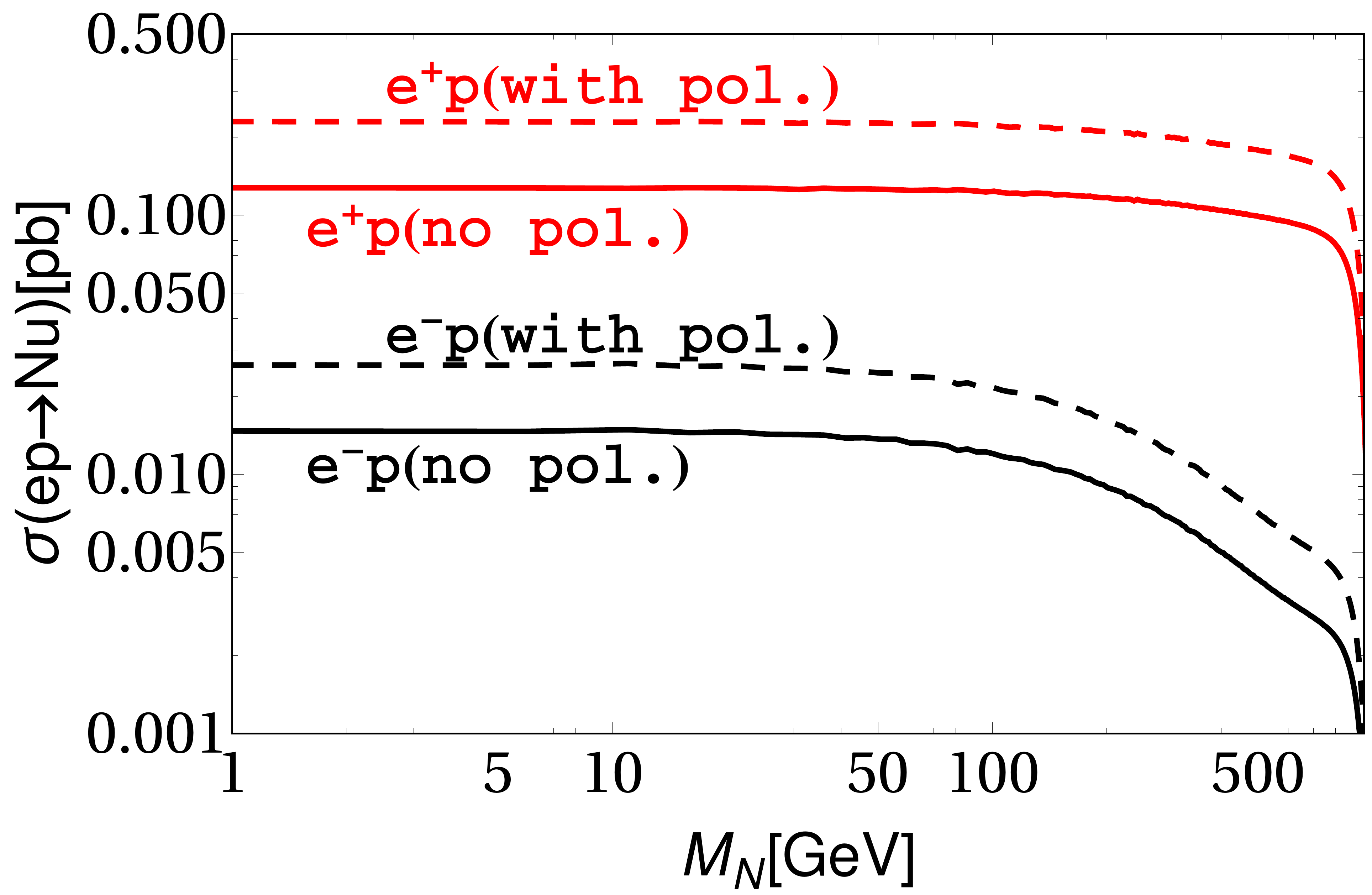}
	\caption{\small{Production cross-section of $ep \to N u $ with and without $80\%$ left~(right) polarised  $e^-(e^+)$ beam.}}  
	\label{fig:cross-section}
\end{figure}
\def\big{\includegraphics[width=0.45\textwidth, height=0.225\textheight]{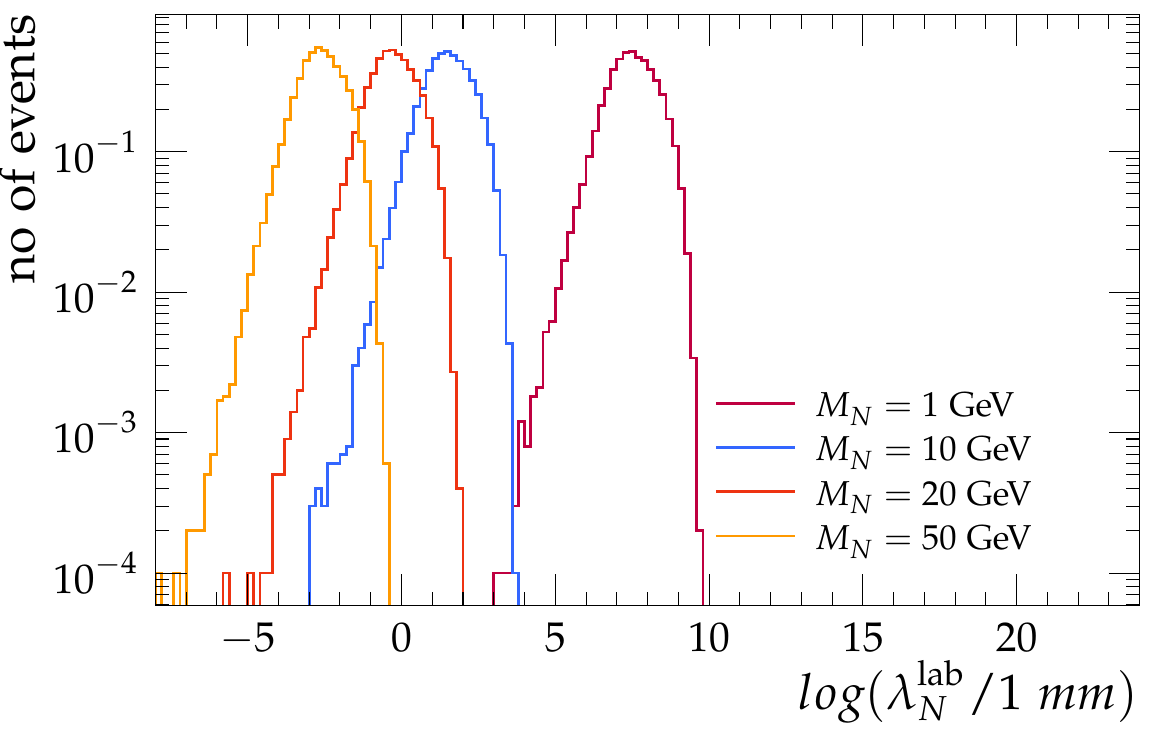}}
\def\little{\includegraphics[width=0.17\textwidth,height=0.1\textheight]{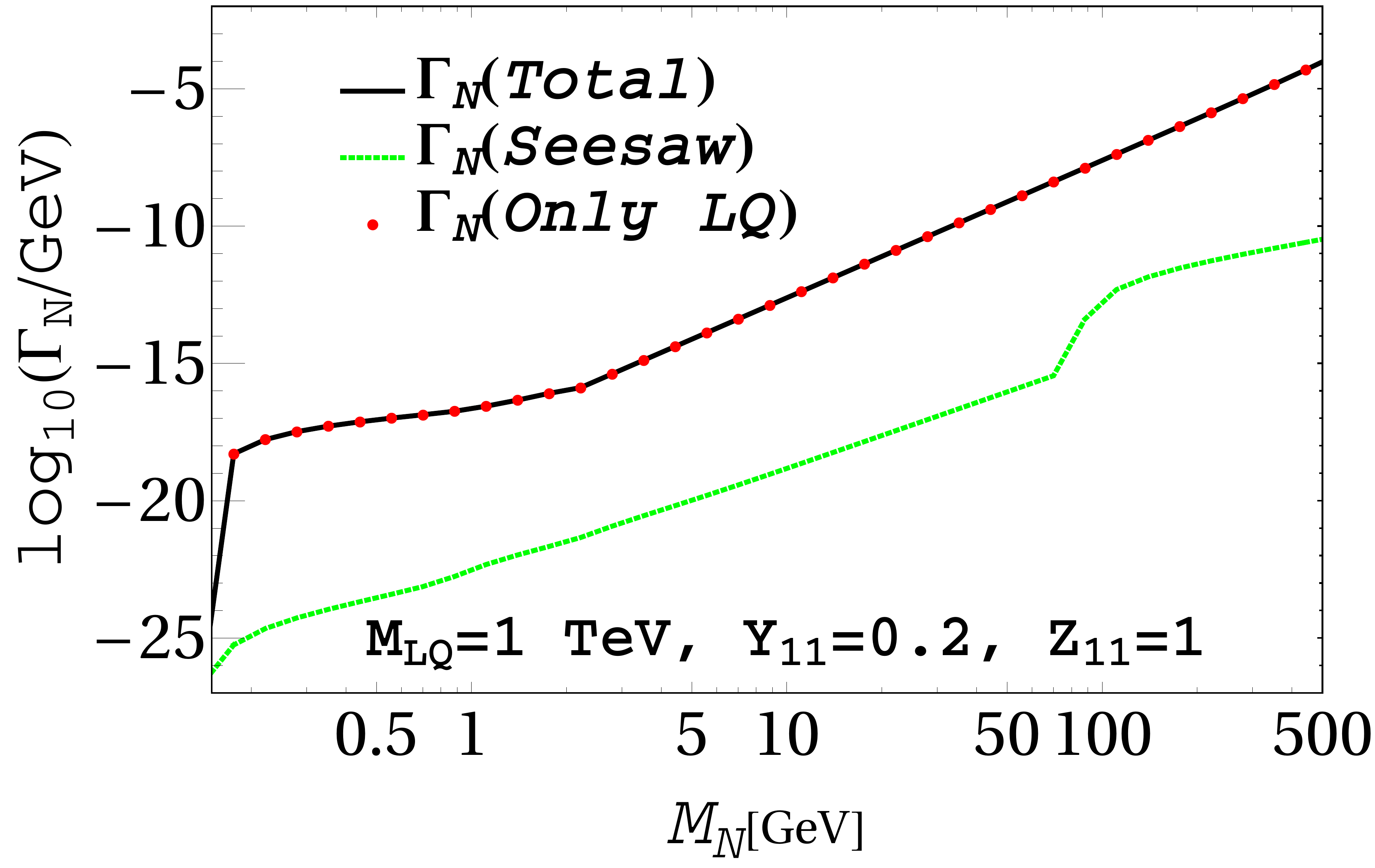}}
\begin{figure}
\centering
\stackinset{l}{135pt}{t}{10pt}{\little}{\big}
\caption{\small{
Distribution of  $\lambda^{\textrm{lab}}_N$, where the  inset figure represents the proper decay width $\Gamma_N$ of $N$. The green and red lines indicate contribution of ($W,Z, h$), and LQ  states in three-body decay of $N$  with the latter dominating $\Gamma_N$. For $W,Z,h$ mediation, which depend on active-sterile mixing, we use 
$m_{\nu}=0.1$ eV.}} 
\label{fig:lifetime}
\end{figure}

We consider both, $e^-$ and $e^+$  in the initial state, \crd{which lead} to different production cross sections \cite{Padhan:2019dcp} due to a difference in the quark PDFs. 
This could be a useful handle to fingerprint the signature, if it were to be observed.
The production cross sections are shown in Fig.~\ref{fig:cross-section}. 
The cross-sections are fairly large, $\sigma \sim  14, 127 $ fb for $e^-$ and $e^+$ beam, respectively,  and increases even further if polarisation of $e^-/e^+$ beam is being used.

For the considered LQ mass $M_{LQ}=1$ TeV, and $N$ with mass $M_N\sim \mathcal{O}(10)$ GeV, $N$ is produced with a boost   and its proper decay length in the laboratory system is enhanced to $\lambda^{\rm lab}_N = \beta \gamma c \tau_N$, where  $\beta \gamma = |p_N|/M_N > 1$ due to RHN momentum $p_N \sim M_{LQ}/2$. 
We show the distribution of $\lambda^{\rm lab}_N$ in Fig.~\ref{fig:lifetime}. As can be seen, $N$ with masses 10, 20 GeV undergoes displaced decays with decay length in the $\mathcal{O}(mm-100 mm)$ range, while for 50 GeV, this  is almost a prompt decay.

\begin{figure}[t]
	\centering	
	\includegraphics[width=0.40\textwidth,height=0.2\textheight]{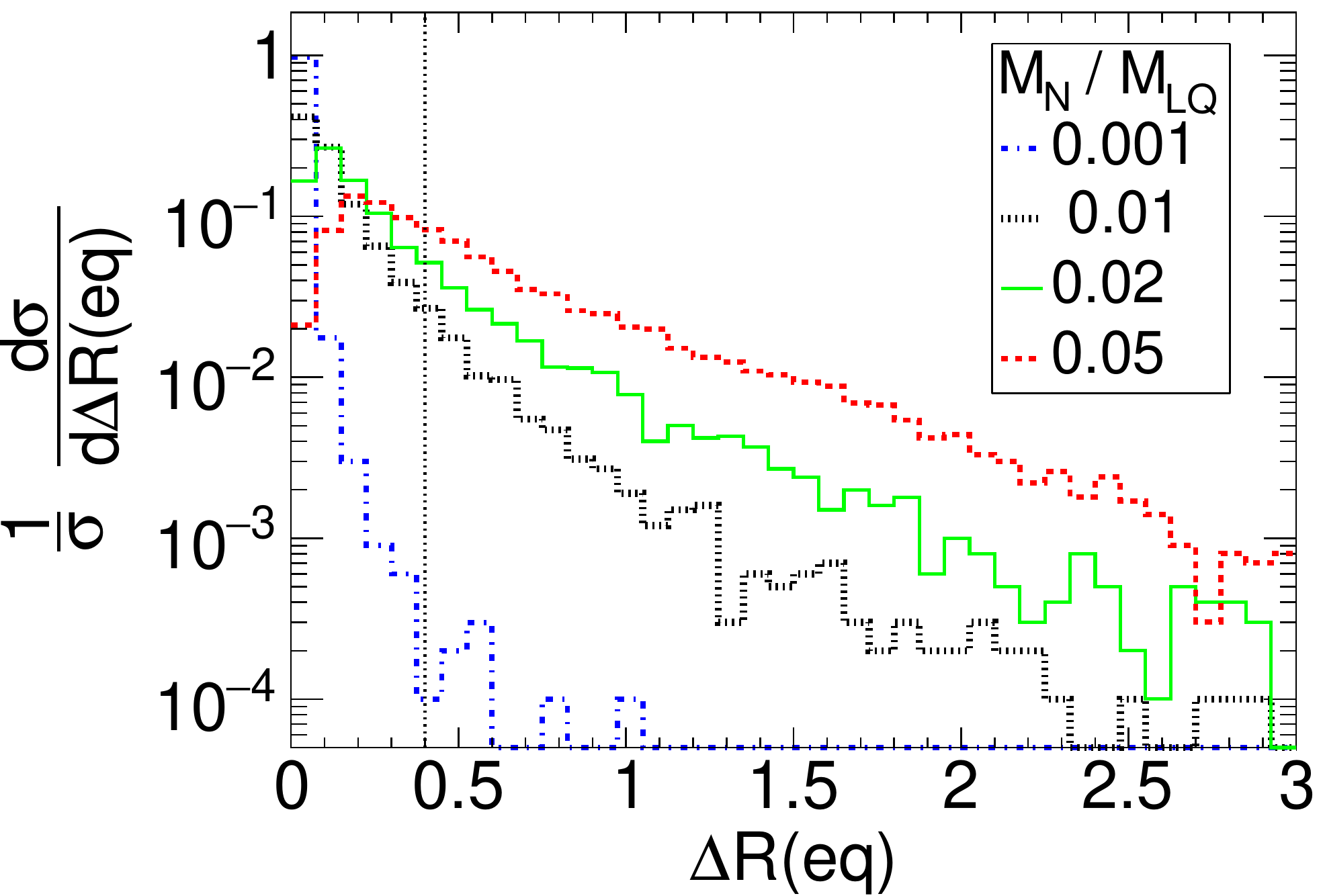}
	\caption{ $\Delta \text{R}(eq)$ separation of closely spaced lepton-quark pair. The vertical line is the typical isolation criterion.} 
	\label{fig:taudeltaR}
\end{figure}

The boost of $N$ also leads to very small angular separation of its decay products, as shown in Fig.~\ref{fig:taudeltaR}. 
Of particular interest is the separation between the charged lepton and the jets. As the figure shows, a sizeable fraction of the decays fails to satisfy the standard lepton isolation criterion, $\Delta R(\ell,j) > 0.4$~($\Delta R$ between other decay products of $N$ also display similar features), which is more pronounced for smaller $M_N$ and implies that the leptons are not sufficiently isolated to be recognized as such. Instead, the $N$ decay products tend to appear as a single jet with  somewhat large radius, referred as a ``fat jet''. Thus, the very specific signature under investigation is a fat jet that originates from a RHN displaced vertex and is accompanied by a prompt jet:
\begin{eqnarray}
e^{\pm} p \to j N \to j + j_{N}^{\rm{displaced-fat}}
\label{eq:eqsig}
\end{eqnarray}
where $j_{N}^{\rm{displaced-fat}}$ denotes the displaced and collimated decay products of the RHN, forming a fat jet, \crd{which we refer as a neutrino jet}. Among the $N \to eq\bar{q}^{\prime}, \nu q \bar{q}^{\prime}$ decay modes, we focus on $N\to e^\pm q \bar{q}^{\prime}$ and include these states in the fat jet description. In terms of sensitivity our results are thus conservative. The chosen decay mode has the added benefit that it allows the reconstruction of $M_N$. Additional channels $N\to \nu q \bar{q}^{\prime}$ can be vetoed by imposing a  missing transverse momentum cut. For a prompt fat jet signature  in inverse seesaw model at LHeC, see \cite{Das:2018usr}. 

\crd{Notice that,  $N$ is produced together with   a prompt  jet with substantial $p_T$ and moderate $\eta$ values. }This jet emerges from the primary vertex and can be used to determine its three-coordinates. The corresponding uncertainty of this determination is related to the tracking precision, which is ${\cal O}(10) \ \mu$m \cite{Agostini:2020fmq,Cheung:2020ndx}. In the subsequent discussion, while designing the cuts, we take for the transverse displacement a detection threshold of 50 $\mu m$, to be conservative.

\begin{figure*}
	\includegraphics[width=0.300\textwidth, height=0.2024\textheight]{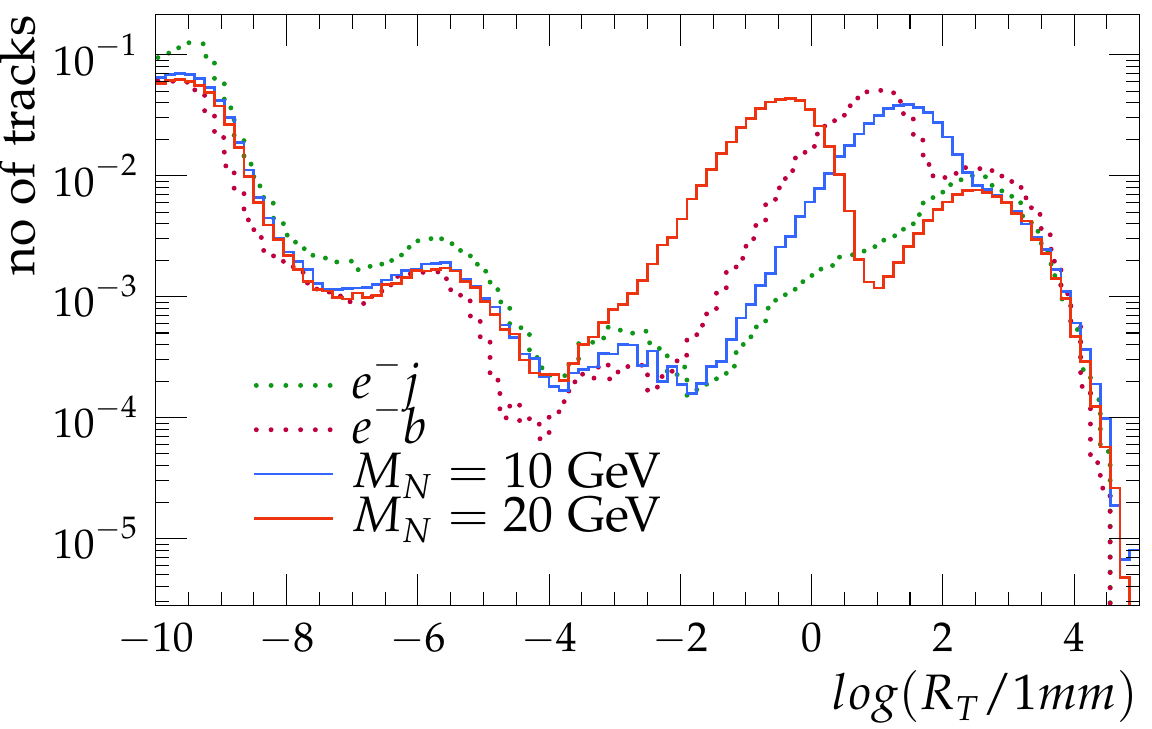}
	\includegraphics[width=0.30\textwidth, height=0.2024\textheight]{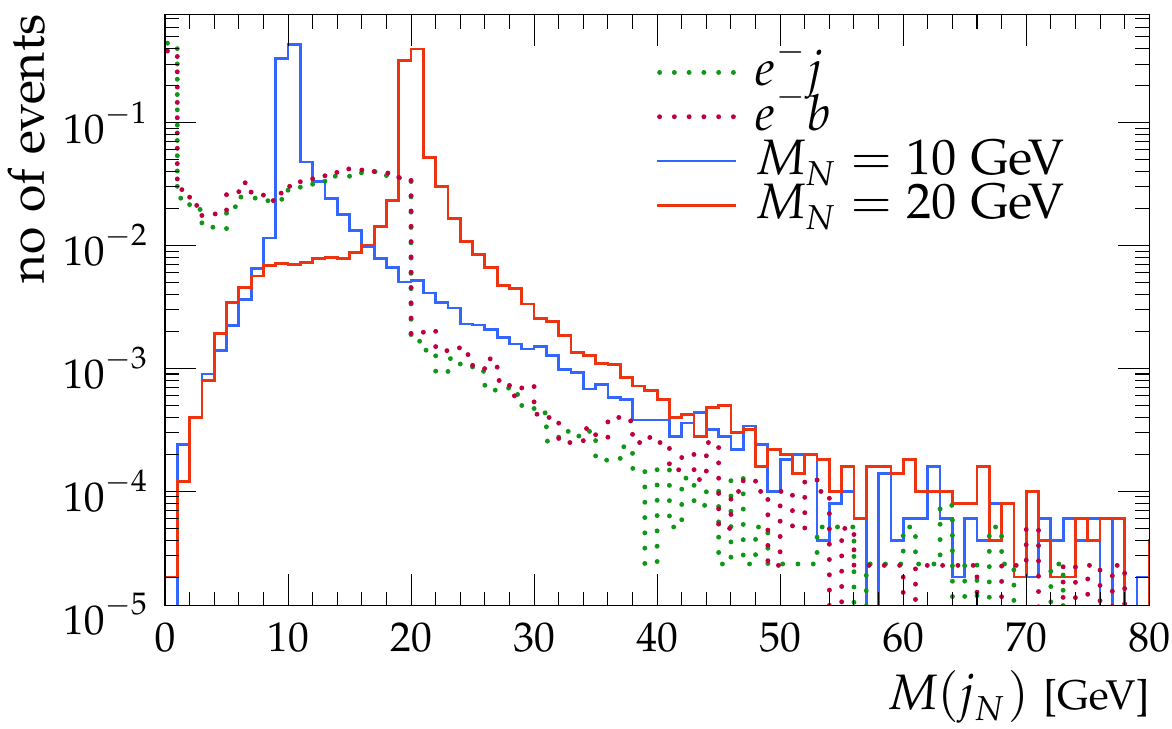}
	\includegraphics[width=0.30\textwidth,height=0.2024\textheight]{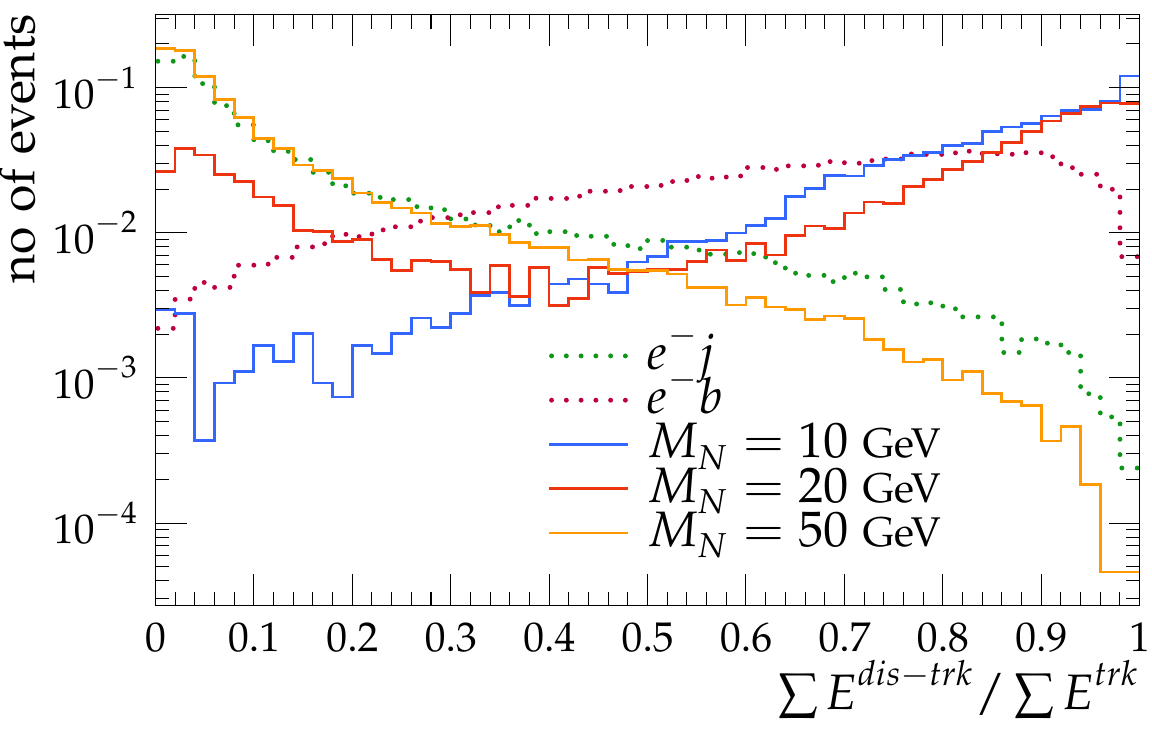}
	\caption{\small{Left panel:  $R_T$ distribution of the  tracks. Middle panel: invariant mass distribution of neutrino jet.  Right panel: distribution of the  summed energy ratio of the displaced tracks and all possible tracks associated to the displaced jet.}} 
	\label{fig:comparison1}
\end{figure*}

{\bf Backgrounds.--} We consider a number of SM backgrounds, $ep \to ej, eb, \nu j, \nu j j, e b j, e b b b$, shown  in Table.~\ref{table:electablea}.  The relevant processes are those involving $B$-hadrons, which can give rise to displaced vertices. The light jet background is also important due to its huge cross-section, see Table.~\ref{table:electablea}. This is especially relevant for the heavier $N$ which are quasi-prompt due to the short lifetime. 
The  single top production $e p \to \nu \bar t$  is already included. Other background involving tau-lepton production $ep \to \nu j (W \to \nu \tau )$ ($\sigma_\tau = 0.01$ pb with our cuts) can be neglected.
 Our main focus is thus  on the final states: $e/\nu+ n_j j+ n_b b$, with $n_j\geq 1$ and/or $n_b \geq 1$ being the number of light quarks and $b$ quarks, respectively. 

\begin{table*}
	\begin{tabular}{|l|l|l|l|l|l|l|l|l|l|l|}\hline
		\multirow{2}{*}{Signal} &  \multicolumn{10}{|l|}{ \hspace{2cm}Background processes $e^-p/e^+p \to abc$}  \\ \cline{2-11}
		\multirow{2}{*}{$ep\to j+j_N$ } & \multirow{2}{*}{$\sigma_b^i$}&$ej$&$eb$  &$\nu j$   & $\nu jj$ & $e bj$&$e jj$&$ebb$&$e+3j_b$&$\nu+ 3j_b$   \\ \cline{3-11} 
		&&6881.8  & 89.9  & 9621.0 & 3482.3  & 20.08 &  2559.4 &32.67 &1238.8  & 1394.3 \\ & & (5099.2) &  (89.99) & (3764.3) &  (1207.0) &  (20.07) & (1787.0) & (32.65) & (890.5) & (486.2) \\ \hline
		 $M_{N}=10$ GeV&&&&&&&&&&\\ 
		$\sigma_s^i=7.37$, $\sigma_s^f=3.5$ ($e^-p$)& $\sigma_b^f$&4.67 &2.86  & 0& 0.42  &0.36 &1.68 &1.04&1.73  &0.47
		  \\ $\sigma_s^i=64.03$, $\sigma_s^f=42.97$ ($e^+p$) &  & (3.46)& (2.87)&   (0)& (0.14)& (0.37) &(1.18)& (1.04)& (1.25)&(0.16) \\ \hline
		$M_{N}=20$ GeV&&&&&&&&&&\\ 
		$\sigma_s^i=7.21$, $\sigma_s^f=3.37$ ($e^-p$) & $\sigma_b^f$&5.09&6.69&0&0.76&1.01&2.51&2.48&3.24&0.72\\
		$\sigma_s^i=63.92$, $\sigma_s^f= 44.65$ ($e^+p$) & &  (3.77) &   (6.69) & (0) & (0.26) & (1.03) &(1.75) & (2.48) & (2.33) & (0.25) \\ \hline
		$M_{N}=30$ GeV&&&&&&&&&&\\ 
		$\sigma_s^i=7.1$, $\sigma_s^f=1.59$ ($e^-p$) & $\sigma_b^f$&1.65&2.83&0.19& 0.42& 0.64&  1.64 & 1.14& 2.0& 0.69 \\
				
				$\sigma_s^i=63.7$, $\sigma_s^f= 25.09$ ($e^+p$) & & (1.22)& (2.84)&(0.07)&(0.14)& (0.65)& (1.14)& (1.14)& (1.44)&(0.24) \\
\hline
		
	\end{tabular}\caption{Initial and after cut cross-sections (in fb) for signal and backgrounds, which are denoted as $\sigma^{i,f}_{s}$ and $\sigma^{i,f}_b$ respectively. For background processes $j_b$ implies including light jet and $b$ jet. The numbers without (within) brackets are for $e^- (e^+)$ mode, respectively.  \label{table:electablea} }
\end{table*}

{\bf Simulation and event selection.--}  
We use MadGraph5 aMC@NLO(v2.7)~\cite{Alwall:2014hca} to simulate both signal and background samples. For the generation of parton level signal events we implement the LQ model in FeynRules(v2.3)~\cite{Alloul:2013bka} and use the UFO files in MadGraph5 aMC@NLO(v2.7). We implement the following generation level cuts for background event $p_{T}(b/j)>20$ GeV, $p_T(\ell)>10$ GeV, $|\eta (b/j/\ell)|<5$, $\Delta R(jj/\ell j/bb/bj/\ell b) > 0.05$. We also demand transverse momentum $p_T$ of leading parton  $> 150$ GeV for background events, so that the majority of  events populate the signal region.  
	We use Herwig (7.2)~\cite{Bellm:2015jjp,Bahr:2008pv} to simulate the hadronization and showering of parton level events. For the signal, we consider the RHN decay in  Herwig (7.2). We use Rivet (v3.0)~\cite{Bierlich:2019rhm,Buckley:2010ar} for event analysis. For jet formation we use  FastJet (v3.3.2)~\cite{Cacciari:2011ma}.
	We reconstruct fat jet using a Cambridge-Aachen algorithm~\cite{Dokshitzer:1997in} with radius parameter $R=1.0$.
 We select events according to the following criteria:
\begin{itemize}
	\item  $N_{jet}\ge2$ and $p_T \ge 50$ GeV for all jets. 
	\item We select all the charged final state particles (tracks) with $p_T>1 $ GeV in an event and calculate the transverse displacement $R_T = \sqrt{X^2 + Y^2}$ ($X,Y$ being the coordinates of the production vertex) of each of the tracks from the interaction point (IP). The distribution is shown in 
	the left panel of Fig.~\ref{fig:comparison1}. For the signal the peaks occurring at a higher value of $R_T$  are due to tracks originating from RHN decay. The high track multiplicity at a lower $R_T$ is due to  tracks associated with the prompt jet. For the light jet background~(also for signal, and $b$ jet background), peak at a higher $R_T~( \simeq 200 \ mm)$ occurs due to the presence of long-lived hadrons.
	
	\item 
	We 
	define the track as a displaced track if the  transverse displacement is above the detection threshold, $R_T>50\, \mu$m. 
	Subsequently, we define  a ratio 
	\begin{eqnarray}
	r_N=\frac{N_{trk}(displaced)}{N_{total-trk}}, 
	\end{eqnarray}
	where ${N_{trk}(displaced)}$ and ${N_{total-trk}}$ are the 
	number of displaced tracks, and total number of tracks  associated to a jet, respectively. 
	Since $N$ is long-lived this ratio is expected to be closer to 1 for jets originating from its decay vertex, while we expect a value $\ll 1$ for any other prompt jet.
	We also impose a bound $R_T<312 \ mm$ , so that we  consider only those decay products appearing  within the tracker volume. We label the jet having the largest value of $r_N$ as the displaced jet. For $N$ with $M_N \sim 50$ GeV the decays are rather prompt, and $r_N$ can not be reliably used to identify the jet as stemming from $N$ decay.

	\item
	We further cross-check if the displaced jet is originating from $N$ \crd{and hence is a neutrino jet} 
	by computing the jet-mass. We identify  a jet as the  neutrino jet, if  its invariant mass $M(j_N)$ is closest to $M_N$. We find that for $M_N \sim 10, 20$ GeV, the jet with highest $r_N$ is consistent this  invariant mass condition. 

	\item We define a variable $r_E$ as the ratio of the sum of energies of the displaced tracks and sum of energies carried out by all tracks associated to a jet:
	\begin{eqnarray}
	r_E=\frac{\Sigma E(displaced-trk)}{\Sigma E(trk)}, 
	\end{eqnarray}
	We show the distribution of this variable in Fig.~\ref{fig:comparison1} for the jet with highest $r_N$. As can be seen, the distribution for displaced signal and light jet background are complementary in nature, which is instrumental in separating the two.
	For small $N$ masses  majority of the tracks are displaced, leading to a higher energy ratio $r_E$. For light jet background, reverse is true.

	\item Finally we select those events where the neutrino jet  satisfies the following cuts: $r_N \ge0.5$, $r_E \ge 0.5$, $ p_T(j_N) \ge 150$ GeV  and $M({j_N})= M_N\pm3$. 
\end{itemize}

\begin{table}[ht]
	\begin{tabular}{|c|c|c|c|} \hline
		$M_{N}$ [GeV] & $n_{\sigma}$ & $\mathcal{L}$ [$\rm{fb}^{-1}$] & {$\mathcal{Y}^{ex}$}
		\\ \hline	
		10 & 6.0 (41.5)  & 34.0 (0.7 ) & {0.067 (0.035)} \\  \hline
		20 & 4.7 (39.7)& 56.8 (0.8) & {0.059} (0.017)	 \\  \hline
		30 & 3.3 (30.4) & 116.6 (1.3) & {0.047} (0.013) \\ \hline
	\end{tabular}
	\caption{ $n_{\sigma}$ is the significance of the proposed 
	signature with  only {$50\,  \textrm{fb}^{-1}$} luminosity. $\mathcal{L}$ is the required luminosity to achieve $5\sigma$ significance. Numbers without (within) brackets corresponds to  $e^-$ ($e^+$) beam, respectively.
	$\mathcal{Y}^{ex}$ represents  2$\sigma$ exclusion on $Y_{11}$.}	
	\label{tab:result}
\end{table}

{\bf{Results.}}- The signal cross-sections for 10-30 GeV $M_N$ varies as  $\sigma^i_s \sim 7.10-7.37$ fb before applying any cut for $e^-$ mode (see Table.~\ref{table:electablea}). The final cross-sections after all the cuts are $\sigma^f_s  \sim 1.6-3.5$ fb.  \crd{For $M_N=50$ GeV as the decay is almost a prompt decay, we do not include the result in the table}. 
We find the backgrounds $ eb, ej$ are the most relevant ones after cuts, while other backgrounds $ebb/ebbb, ejj/ejjj$ 
also give sizeable contributions. 
Applying a veto on missing transverse momentum will further suppress the $\nu j/\nu jj/\nu+ 3j_b$ backgrounds. Instead, here we present a conservative estimate. 
For the $e^+$ mode, the signal cross-section is one order of magnitude larger as compared to $e^-$ mode, 
 while the backgrounds are relatively smaller. 
The signal sensitivity  is calculated as 
\begin{equation}
n_{\sigma}=\sqrt{\mathcal{L}} \frac{\sigma^f_s}{\sqrt{\sigma^f_s+\sigma^f_b}}
\label{eqsigma}
\end{equation}
where $\mathcal{L}$ in $\rm{fb}^{-1}$ is the required luminosity to achieve $n_{\sigma}$ significance, $\sigma^f_{s}/\sigma^f_b$ are the after-cut signal/background cross sections. Our findings are summarised in Table~\ref{tab:result}. 

For the considered RHN masses,  our signature can be detected with $ 5\sigma $ significance with an integrated luminosity $\mathcal{L}<120$\, $\rm{fb}^{-1}$. 
With  the total integrated luminosity of 1000 $\rm{fb}^{-1}$ in the $e^-$ mode, excluding our signal at 2$\sigma$ significance can be used to put an upper limit on Yukawa $Y_{11}\leq 0.067$ for $M_N=10$ GeV. 
For higher RHN mass $M_N= 30$ GeV,  the limit becomes stringent  due to the increase in geometric cut-efficiency ($R_T<312 \ mm$). 

The table further shows that the $e^+$ mode clearly outperforms the $e^-$ mode as signal sensitivity $n_{\sigma}$ increases by almost one order of magnitude, and  $5\sigma$ significance  can be achieved with luminosities $\mathcal{L}<2\, \rm{fb}^{-1}$ only, for all the mass points considered. 
With $\mathcal{L}=1000\, \textrm{fb}^{-1}$ the upper limit on Yukawa tightens   $Y_{11} \leq $0.013 for $M_N=30$ GeV.
The  signal sensitivity can be further improved   with  polarised $e^-$ and  $e^+$ beams due to increase in signal cross-section. 

{\bf Conclusions.--}
In this article we considered a model that includes an $\tilde R_2$ LQ with $M_{LQ}=1$ TeV and  RHN with a mass of ${\cal O}(10)$ GeV. The RHN can be produced from resonant LQ decay, with comparatively large cross sections, and also via $t$ channel processes. 
We studied the prospects of discovering the RHN at the LHeC via a displaced fat jet signature, which is purely hadronic in nature and therefore difficult to detect at the LHC. We perform a detailed analysis of the signal and the different SM background processes. We find that the ratio of energy deposits of the displaced and all possible tracks associated with the displaced jet is instrumental in separating displaced  decays of RHN from background. 
We find that RHN in the considered mass range can be detected at the LHeC with only \crd{$\mathcal{L}<120 \ (2)\, \rm{fb}^{-1}$ luminosity with $e^- (e^+)$ beam}. 
\crd{We observe that, the use of a positron beam at LHeC clearly  enhances  the detection prospect of this signature  by order of magnitude. }


\medskip
\subsection*{Acknowledgements}
  G.C. acknowledges support from ANID FONDECYT-Chile grant No. 3190051.
  The work of S.M. is supported by the Spanish grant FPA2017-85216-P (AEI/FEDER, UE), PROMETEO/2018/165 (Generalitat Valenciana). M.M thanks Indo-French Centre for the Promotion of Advanced Research for the funding (project no: 6304-2). 

\bibliography{bibitem}
\bibliographystyle{apsrev4-1}

\end{document}